\documentclass[sigplan,screen]{acmart}

\usepackage{graphicx}
\usepackage{enumitem}
\usepackage{amsmath}
\usepackage{caption}
\usepackage{subcaption}
\usepackage{multirow}
\usepackage{hyperref}
\usepackage{fancyvrb}
\usepackage{xcolor}
\usepackage{siunitx}
\usepackage{microtype}

\newcommand{\red}{\textcolor{red}}
\newcommand{\blue}{\textcolor{blue}}



\title{Efficient GPU Implementation of Affine Index Permutations on Arrays}
\author{Mathis Bouverot-Dupuis} 
\affiliation{%
  \institution{ENS Paris}
  \city{Paris}
  \country{France}
}
\email{mathis.bouverot@ens.psl.eu}
\author{Mary Sheeran} 
\affiliation{%
  \institution{Chalmers University}
  \city{Gothenburg}
  \country{Sweden}
}
\email{mary.sheeran@chalmers.se}
\date{May 2023}

\setcopyright{acmlicensed}
\acmPrice{15.00}
\acmDOI{10.1145/3609024.3609411}
\acmYear{2023}
\copyrightyear{2023}
\acmSubmissionID{icfpws23fhpncmain-id2-p}
\acmISBN{979-8-4007-0296-9/23/09}
\acmConference[FHPNC '23]{Proceedings of the 11th ACM SIGPLAN International Workshop on Functional High-Performance and Numerical Computing}{September 4, 2023}{Seattle, WA, USA}
\acmBooktitle{Proceedings of the 11th ACM SIGPLAN International Workshop on Functional High-Performance and Numerical Computing (FHPNC '23), September 4, 2023, Seattle, WA, USA}
\received{2023-05-31}
\received[accepted]{2023-06-28}

\begin{document}

\begin{abstract}
Optimal usage of the memory system is a key element of fast GPU algorithms. Unfortunately many common algorithms fail in this regard despite exhibiting great regularity in memory access patterns. In this paper we propose efficient kernels to permute the elements of an array. We handle a class of permutations known as Bit Matrix Multiply Complement (BMMC) permutations, for which we design kernels of speed comparable to that of a simple array copy. This is a first step towards implementing a set of array combinators based on these permutations.
\end{abstract}

\maketitle

\section{Introduction}
\label{sec:intro}

In GPU algorithms, memory access is often the performance bottleneck. Consider the following low-level GPU kernel that transforms an array of size $2^n$~:
\begin{Small}
\begin{verbatim}
kernel(int* input, int* output) 
{
  i <- thread_id();
  x <- input[bit-rev(n, i)];
  ...do...stuff...
  output[i] <- y;
}
\end{verbatim}
\end{Small}

Here, $2^n$ threads are launched that each read and write a single element. The read position is computed using the \texttt{bit-rev} function that performs bit-reversal on the index $i$ viewed as a list of $n$ bits, so that $\texttt{bit-rev}(4, 7)$ transforms $7 = 0\texttt{b}0111$ to $14 = 0\texttt{b}1110$. 

Behind this deceivingly simple access pattern lies terrible performance; the read is typically an order of magnitude slower than the write on modern GPUs. Despite the great degree of regularity present in this memory access pattern, it yields uncoalesced memory accesses that force the reads from different threads to be serialized.

While bit reversal often has a hardware or low level implementation, many other transformations (such as those in sorting networks) exhibit a similar degree of regularity that is not fully exploited by GPUs. To this end, we use an alternative way of describing array indexing that allows many regular access patterns to be compiled to efficient GPU code. We view indices into an array of size $2^n$ as binary vectors of size $n$ (vectors in $F_2^n$) and focus on affine transformations in $F_2^n$, the so-called Binary Matrix Multiply and Complement (BMMC) transformations ~\cite{cormen-bmmc-implementation, edelman-index-transformations}. We study the BMMC permutations because they enable reasoning about and implementation of the sets of combinators that we have earlier considered for both software and hardware design~\cite{obsidian-paper, lava-sorter}.

The contributions of this paper are as follows~:
\begin{itemize}
  \item We show how to efficiently implement a specific class of array permutations where the mapping between indices is given by a BMMC\footnote{The code to generate and benchmark our CUDA kernels is publicly available at \url{https://github.com/MathisBD/bmmc-perms-gpu}.}. 
  \item We conduct an empirical evaluation of our kernels, both in the worst case and the average case. 
  \item We show preliminary work on using BMMC permutations to compile high level array combinators.
\end{itemize} 

More precisely, we show how to implement a subclass of BMMC permutations - namely \textit{tiled} BMMC permutations - almost as fast as a simple array copy, and how to factorize any BMMC as the product of at most two tiled BMMCs.

\section{Background~: GPU Programming}
\label{sec:background}

\subsection{A simple GPU model}

This section presents the relevant parts of a simple GPU model that we will use to justify our optimizations. There are two key aspects to this model~: the execution model and the memory hierarchy. For a more in depth discussion of a similar machine model we refer the reader to chapter $4$ "Parallelism and Hardware Constraints" of Henriksen's thesis on Futhark~\cite{futhark-thesis}.

Regarding terminology, there are unfortunately two distinct sets of terms; we will be using the CUDA set, which differs from but also overlaps with the OpenCL set.

The execution model follows an SIMT (single instruction multiple threads) design; a large number of threads are launched concurrently, all executing the same code. Threads are uniquely identified by a thread identifier, which often dictates how they will behave. They are organized according to the following hierarchy~:
\begin{itemize}[label={},itemindent=-2em,leftmargin=2em]
  \item \textbf{Kernels} are the top-level scheduling unit~: all threads in a kernel execute the same code. To obtain good performance it is necessary that a kernel have many threads (typically at least a 100 thousand), and in general there is no kernel-level synchronization possible between threads. A GPU program consists of one or several kernels that are run sequentially. 
  \item \textbf{Thread blocks} are the unit at which thread synchronization - whether it be memory or execution synchronization - can happen. In kernels where the threads do not need synchronization (map-like kernels), the thread group is mostly irrelevant. Maximum thread block size is hardware dependent~: typical sizes are 256 and 1024 threads for AMD and NVIDIA GPUs respectively.
  \item \textbf{Warps} form the basic unit for execution and scheduling. Threads inside a single warp execute instructions in lockstep, including memory access instructions, so that all memory transactions of a warp must have completed before it can advance to the next instruction. Warp size is hardware dependent, although 32 threads is typical.
\end{itemize}

Kernels usually launch many more threads than can be run concurrently. In this case, threads are launched one thread block at a time, with new thread blocks being swapped in as previous blocks finish execution. The order in which blocks are scheduled is by increasing thread identifier~: this means that at any given time the threads currently in flight cover a contiguous subset of the thread identifiers.

The other side of the coin is the GPU memory hierarchy, which reflects the thread hierarchy~:
\begin{itemize}[label={},itemindent=-2em,leftmargin=2em]
  \item \textbf{Global memory} is large off-chip memory (typically on the order of several GiB). This is where the CPU copies data to and from, and is where the inputs and outputs to a kernel reside. If accessed properly global memory has a much larger bandwidth than usual CPU RAM.
  \item \textbf{Shared memory} is smaller and shared by all threads in a thread group. It usually functions as a cache used by thread blocks~: however unlike traditional caches, the programmer is responsible for loading data in and out of shared memory.
  \item \textbf{Registers} are small bits of memory private to each thread. Although very fast, the number of registers per thread is limited. Kernels that require many registers per thread will cause fewer threads to run concurrently. 
\end{itemize}

\subsection{Optimizing memory access}

In contrast to CPUs, GPU programmers must manually manage most of the memory hierarchy in order to get the best performance. Hardware managed caches, while also present on GPUs, are of less importance; most performance benefits come from mechanisms that allow certain memory transactions to be answered concurrently, known as \textit{coalesced} and \textit{bank conflict free} memory accesses.

\begin{figure}[htpb]
  \centering
  \begin{subfigure}[b]{0.4\textwidth}
    \centering
    \includegraphics[width=\textwidth]{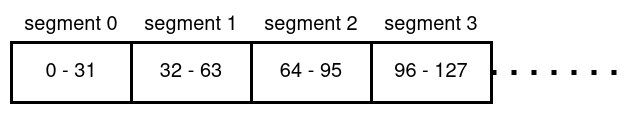}
    \caption{Global memory layout assuming each segment is of size $128$ bytes.}
    \label{fig:global-mem-layout}
  \end{subfigure}
  \begin{subfigure}[b]{0.35\textwidth}
    \centering
    \includegraphics[width=\textwidth]{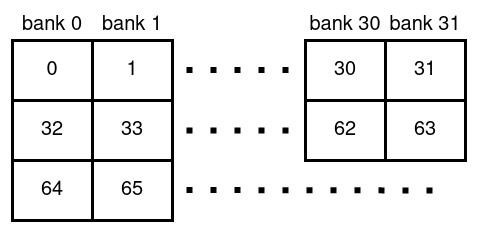}
    \caption{Shared memory layout.}
    \label{fig:shared-mem-layout}
  \end{subfigure}
  \caption{Typical layouts for global and shared memory. The numbers correspond to addresses of 32-bit words.}
\end{figure}

Global memory is divided into contiguous segments - typically 32, 64 or 128 bytes - that form the basic unit for memory transactions (see Figure \ref{fig:global-mem-layout}). The size of a segment is much larger than what can be accessed by a single thread in a given instruction, and in general the memory transactions needed for the individual threads in a warp are serialized. However modern GPUs ensure that the memory accesses from a given warp that fall in the same segment are \textit{coalesced} into one transaction (the order of addresses within a segment does not matter). To obtain optimal memory performance the set of segments accessed by a warp must be as small as possible. Memory access patterns that fail to exploit coalescing can lead to over an order of magnitude decrease in bandwidth.

Shared memory is divided into \textit{banks} (typically 32). Contrary to global memory segments, shared memory banks are not contiguous but rather interleaved at the 32-bit word granularity~: see Figure \ref{fig:shared-mem-layout} for an illustration. Accesses by a warp that fall in the same memory bank must be serialized, but accesses to different banks can be answered concurrently. If threads within a warp access the memory banks in an imbalanced way, a bank conflict occurs, potentially causing a decrease in shared memory bandwidth of up to 32 times.

\subsection{An example~: matrix transposition}
\label{sec:background-transpose}

To help gain some intuition on GPU programming we walk through an example kernel. Let $M$ be a two-dimensional matrix of size $(N, N)$. We would like to write a kernel that performs matrix transposition on $M$~: $M[i, j] \leftarrow M[j, i]$ for all $i$ and $j$, and $M$ is stored in row major order in both the input and output.

If we assume that $N$ is a power of two, we can write the following kernel (in CUDA-like pseudocode)~: 

\begin{Small}
\begin{verbatim}
kernel transpose_naive(int* input, int* output)
{
  size_t i = blockIdx.y * blockDim.y + threadIdx.y;
  size_t j = blockIdx.x * blockDim.x + threadIdx.x;

  output[j * N + i] = input[i * N + j];
}
\end{verbatim}
\end{Small}

The variables \texttt{blockIdx}, \texttt{blockDim} and \texttt{threadIdx} are three-dimensional vectors that store, for each thread, the corresponding block index, block size and thread index within its block.

We invoke this kernel with a grid of $(N/32)*(N/32)$ thread blocks with each thread block being of size $32*32$. When using a two-dimensional indexing scheme for thread blocks (as is done here) the index of a thread within its block is given by \texttt{threadIdx.y * blockDim.x + threadIdx.x}, and warps correspond to bundles of 32 threads that have contiguous indices. In this case, each warp corresponds to a single value for \texttt{i} and 32 contiguous values for \texttt{j}. This means that the first memory access (reading the input) is fully coalesced, but the second memory access (writing the output) is not.

\begin{figure}[H]
  \centering
  \begin{subfigure}[b]{0.35\textwidth}
    \centering
    \includegraphics[width=\textwidth]{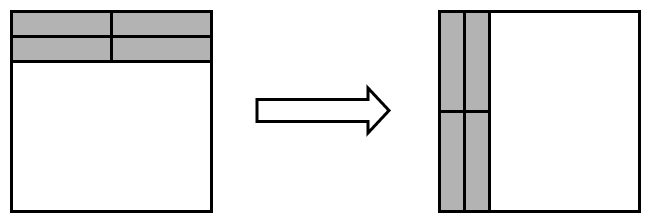}
    \caption{Naive transpose.}
  \end{subfigure}
  \begin{subfigure}[b]{0.35\textwidth}
    \centering
    \includegraphics[width=\textwidth]{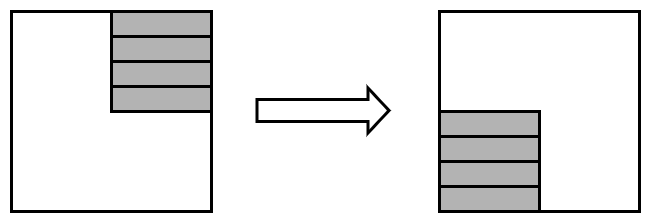}
    \caption{Tiled transpose.}
  \end{subfigure}
  \caption{The shaded area represents the part of matrix $M$ that is accessed by a single thread group, in both input and output. This area is further divided into regions that are accessed by individual warps (for visual clarity, we drew only $4$ warps per thread group; in reality there would be $32$).}
  \label{fig:transpose-patch}
\end{figure}

To ensure that both memory accesses are coalesced we can make use of shared memory. Each thread block will process a square tile of the input of size $32*32$ (compare this to the naive kernel where each block processes a contiguous patch of the input, see figure \ref{fig:transpose-patch} for an illustration). When reading in the tile, each warp will process a single row of the tile, but when writing out the tile, each warp will process a single column of the tile~:

\begin{Small}
\begin{verbatim}
kernel transpose_tiled(int* input, int* output)
{
  shared tile[32*32];
    
  size_t block_i = blockIdx.y * blockDim.y;
  size_t block_j = blockIdx.x * blockDim.x;
  size_t i = threadIdx.y;
  size_t j = threadIdx.x;

  tile[i * 32 + j] = 
      input[(block_i + i) * N + block_j + j];
  synchronize();
  output[(block_j + i) * N + block_i + j] = 
      tile[j * 32 + i];
}
\end{verbatim}
\end{Small}

Each thread group uses an array \texttt{tile} of size $32 * 32$ in shared memory. We have to manually \texttt{synchronize()} threads within a thread block so that the tile for this block is fully populated before we start writing out. The tile processed by a given block has its upper left corner at position $(\texttt{block\_i}, \texttt{block\_j})$ in the input, which corresponds to the tile with upper left corner $(\texttt{block\_j}, \texttt{block\_i})$ in the output.

We measured the performance of the above transpose kernels for matrices of size $2^{15} * 2^{15}$ on an NVIDIA RTX4090 GPU. The effective memory bandwidth achieved in each case is computed by comparing the running time to that of a simple copy kernel~:

\begin{center}
\begin{tabular}{|c|c|c|}
  \hline
  kernel & running time & effective bandwidth \\
  \hline
  copy & \SI{9.3}{ms} & $100\%$ \\
  \hline
  naive transpose & \SI{26.4}{ms} & $35.2\%$ \\
  \hline
  tiled transpose & \SI{12.2}{ms} & $76.2\%$ \\
  \hline
\end{tabular}
\end{center}

The tiled version is over twice as fast as the naive version. Further optimizations can bring the running time even closer to the copy kernel~: we refer the interested reader to the NVIDIA tutorial~\cite{nvidia-transpose-tutorial}.

\section{Key ideas}
\label{sec:key-ideas}

Viewing indices into arrays of size $2^n$ as binary vectors of length $n$ allows us to restrict our attention to certain well-behaved transformations on indices. Arguably the simplest transformations according to this point of view are linear and affine mappings, i.e. mappings between source indices $x$ and target indices $y$ such that~:
$$y = A x + c$$ 
where $A$ is an $(n, n)$ binary matrix, $c$ is a binary vector of length $n$ and all arithmetic is done modulo $2$ (i.e. in $F_2$ the finite field with two elements).
If we expand this formula, each bit of $y$ is given by~:
$$y_i = \left( \sum_{0 \leq j < n} a_{ij} x_j \right) + c_i$$

Many common transformations on indices can in fact be expressed in this way. For instance, transposing a matrix of size $4 * 4$ can be expressed as follows~:
\[
y_i = x_{\texttt{$(i+2)$ \% $4$}}
\hspace{0.5cm} \texttt{i.e.} \hspace{0.5cm}
\begin{bmatrix} 
  y_0 \\ y_1 \\ y_2 \\ y_3 
\end{bmatrix} = 
\begin{bmatrix}
  0 & 0 & 1 & 0 \\
  0 & 0 & 0 & 1 \\
  1 & 0 & 0 & 0 \\
  0 & 1 & 0 & 0
\end{bmatrix}
\begin{bmatrix} 
  x_0 \\ x_1 \\ x_2 \\ x_3 
\end{bmatrix} +
\begin{bmatrix}
  0 \\ 0 \\ 0 \\ 0
\end{bmatrix}
\]
The above matrix has exactly one non-zero entry per row and per column. Invertible matrices of this form are called {\em permutation matrices} and simply permute the bits of the input index. In the above example, the index with bits $[ x_0, x_1, x_2, x_3 ]$ is mapped to $[ x_2, x_3, x_0, x_1 ]$, so that index $6 = 0\texttt{b}0110$ is mapped to index $9 = 0\texttt{b}1001$.

When the matrix $A$ is a permutation matrix and the complement vector $c$ is $0$ we call $(A, c)$ a Bit Permute (BP) transformation. Bit-reversal is thus a BP transformation~:
\[
y_i = x_{n-1-i}
\hspace{0.5cm} \texttt{i.e.} \hspace{0.5cm}
\begin{bmatrix} 
  y_0 \\ y_1 \\ y_2 \\ y_3 
\end{bmatrix} = 
\begin{bmatrix}
  0 & 0 & 0 & 1 \\
  0 & 0 & 1 & 0 \\
  0 & 1 & 0 & 0 \\
  1 & 0 & 0 & 0
\end{bmatrix}
\begin{bmatrix} 
  x_0 \\ x_1 \\ x_2 \\ x_3 
\end{bmatrix} +
\begin{bmatrix}
  0 \\ 0 \\ 0 \\ 0
\end{bmatrix}
\]

The complement vector is also useful. Here is an example of using it to define a transformation that reverses an array of size $16$~:
\[
y_i = x_i + 1
\hspace{0.5cm} \texttt{i.e.} \hspace{0.5cm}
\begin{bmatrix} 
  y_0 \\ y_1 \\ y_2 \\ y_3 
\end{bmatrix} = 
\begin{bmatrix}
  1 & 0 & 0 & 0 \\
  0 & 1 & 0 & 0 \\
  0 & 0 & 1 & 0 \\
  0 & 0 & 0 & 1
\end{bmatrix}
\begin{bmatrix} 
  x_0 \\ x_1 \\ x_2 \\ x_3 
\end{bmatrix} +
\begin{bmatrix}
  1 \\ 1 \\ 1 \\ 1
\end{bmatrix}
\]
In this case the matrix $A$ is also a permutation matrix (corresponding to the identity permutation) but the complement vector $c$ is non-zero~: we call such a transformation a Bit Permute Complement (BPC).

In the most general case, $A$ is any invertible matrix (over $F_2$) and $c$ is any vector, giving a Bit Matrix Multiply Complement (BMMC) transformation. The invertibility requirement for $A$ ensures that the transformation defines a permutation on arrays. While not all permutations on arrays can be expressed in such a way, the preceding examples should convince the reader that this class includes many of the common cases. 
Note that permutations on an array whose size is not a power of 2 do not fall in the BMMC class. For instance transposing a matrix of size $(7, 5)$ is not a BMMC permutation, whereas transposing a matrix of size $(16, 8)$ is.

BMMCs were studied in the context of data-parallel programming in the 1990s by Cormen, Edelman and their co-authors. Both exploited the power of linear algebra, such as various matrix decompositions or Gaussian elimination, inspiring this work~\cite{cormen-bmmc-implementation, edelman-index-transformations}. For example, Cormen proposed asymptotically optimal implementations for BMMC permutations on the disk I/O model~\cite{cormen-bmmc-implementation}.

\begin{figure}[htpb]
    \centering
    \includegraphics[width=0.8\linewidth]{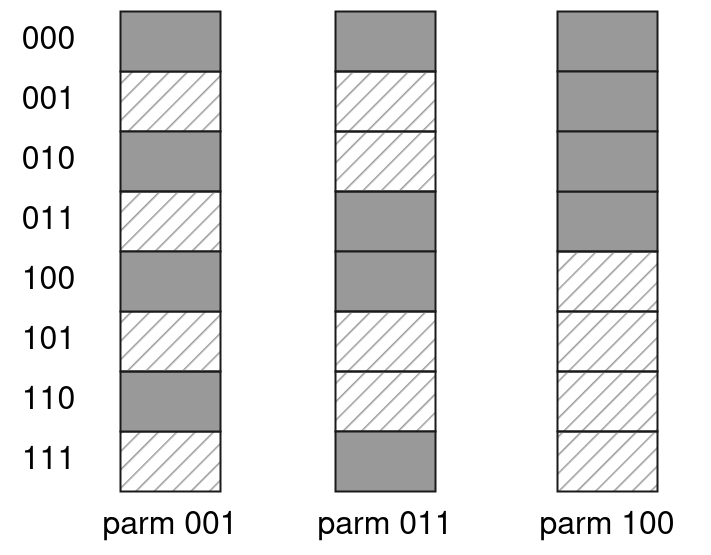}
    \caption{Applying \texttt{parm} to an array of size $8$ with different masks written in binary. The first and second subarrays are represented respectively with a solid and dashed background.}
    \label{fig:parm-examples}
\end{figure}

We aim to use BMMC permutations to provide an efficient implementation for high-level combinators that allow the programmer to describe data access patterns concisely. In this paper, we give one example of such a combinator, called \texttt{parm}. The expression \texttt{parm mask f xs} partitions the array \texttt{xs} of size $2^n$ into two equally sized subarrays according to the $n$ bit \texttt{mask}, applies \texttt{f} to each subarray and stitches the resulting arrays back together according to the \texttt{mask}. The element at index \texttt{i} is assigned to the first or second subarray according to the dot product $\texttt{i} * \texttt{mask}$ in $F_2$~: see Figure \ref{fig:parm-examples} for examples. 

In section \ref{subsec:parm-bmmc}, we show how to efficiently implement the \texttt{parm} combinator in terms of BMMC permutations. In fact, for any mask $m$ we can find a matrix $A$ such that~:
\[
\texttt{parm} \: m \: f = \texttt{bmmc} \: (A^{-1}, 0) \circ \texttt{parm} \: 2^{n-1} \: f \circ \texttt{bmmc} \: (A, 0)
\]
where $\texttt{parm} \: 2^{n-1} \: f$ applies $f$ to the first and second halves of the input array, and composition is from right to left. For instance, the BMMC corresponding to \texttt{parm $0$b$0011$} is
\[
\begin{bmatrix} 
  y_0 \\ y_1 \\ y_2 \\ y_3 
\end{bmatrix} = 
\begin{bmatrix}
  0 & 1 & 0 & 0 \\
  0 & 0 & 1 & 0 \\
  0 & 0 & 0 & 1 \\
  1 & 1 & 0 & 0
\end{bmatrix}
\begin{bmatrix} 
  x_0 \\ x_1 \\ x_2 \\ x_3 
\end{bmatrix} +
\begin{bmatrix}
  0 \\ 0 \\ 0 \\ 0
\end{bmatrix}
\]
We are currently working on implementing \texttt{parm} and related combinators in the Futhark programming language, a high-level functional language that can compile to efficient GPU code~\cite{futhark-pldi}. These combinators benefit greatly from fusion rules such as the following~:
\[
\texttt{bmmc} \: (A, c) \circ \texttt{bmmc} \: (B, d) = \texttt{bmmc} \: (AB, Ad + c) 
\]
Some challenges arise when compiling uses of the bmmc combinator in nested parallel code. In fact for any BMMC $(A, c)$ of size $n$ and any mask $m$ we can find another BMMC $(A', c')$ of size $n+1$ such that~:
\[
\texttt{parm} \: m \: (\texttt{bmmc} \: (A, c)) = \texttt{bmmc} \: (A', c') 
\]

The main contribution of this paper is to give an efficient implementation for a class of BMMC permutations that we call \textit{tiled} BMMC permutations. These include all BPC permutations, such as transpose and bit reverse. We show how to generalize the matrix transposition kernel to tiled BMMCs and compare the impact of different optimizations in sections \ref{sec:bpc} and \ref{sec:results}. The final kernels we obtain are fully coalesced and bank-conflict free, reaching on average $90\%$ of the maximum effective memory bandwidth.  

Finally, we show in section \ref{sec:bmmc} how to use linear algebra techniques to decompose any BMMC $A$ as a product $A = T_1 T_2$ of two tiled BMMCs. The permutation defined by $A$ can then be efficiently realized by first applying the kernel for $T_2$ followed by the kernel for $T_1$.

It should be noted that we assume an offline setting, i.e. that the BMMC matrix and complement vector are known in advance (before generating the CUDA code for the kernel). This is in accordance with our aim to implement the techniques described in this paper in the Futhark compiler.

\section{Implementing BPC permutations}
\label{sec:bpc}

In this section, we explain how to generalize the transpose kernel from section~\ref{sec:background-transpose} to arbitrary BPC permutations. We start by introducing simple tiling to enable coalesced memory access before gradually adding further optimizations. As a running example the reader can inspect the different kernels generated for the bit-reverse permutation in the appendix.

\subsection{Ensuring coalesced accesses}
\label{subsec:bpc-coalesce}

\begin{figure}[ht]
  \centering
  \begin{subfigure}[b]{0.45\textwidth}
    \centering
    \includegraphics[width=\textwidth]{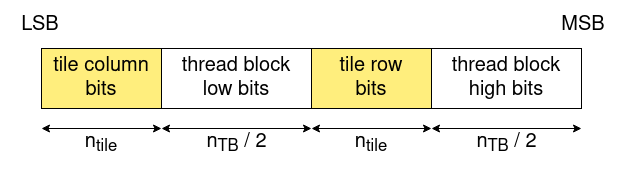}
    \caption{Transpose (for a square matrix).}
  \end{subfigure}
  
  \begin{subfigure}[b]{0.45\textwidth}
    \centering
    \includegraphics[width=\textwidth]{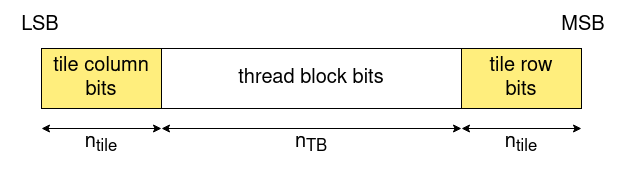}
    \caption{Bit reverse~: $p(i) = n-1-i$}
  \end{subfigure}

  \begin{subfigure}[b]{0.45\textwidth}
    \centering
    \includegraphics[width=\textwidth]{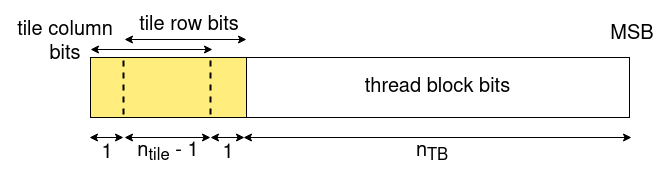}
    \caption{Cyclic shift~: $p(i) = \begin{cases} i-1 & \texttt{ if } i > 0 \\ n-1 & \texttt{ if } i = 0  \end{cases}$}
  \end{subfigure}
  
  \caption{Partition of input index bits for different permutations. In the first two cases $n\_over = 0$, and in the third case $n\_over = n\_tile - 1$. Note that in general the tile row bits need not be contiguous, and so do the thread block bits.}
  \label{fig:bit-partition}
\end{figure}

The first step is to define the notion of tile for an arbitrary BPC $(p, c)$, where $p$ is a permutation on $\{0, \dots, n-1\}$ and $c$ is a complement vector. We start by partitioning the bits of input indices as follows~:
\begin{itemize}
    \item The tile column bits are the $n\_tile$ lower bits.
    \item The tile row bits are the $n\_tile$ bits such that\\$p(bit\_index) < n\_tile$.
    \item The tile overlap bits are the $n\_over$ bits that are both tile column and tile row bits.
    \item The thread block bits are the $n\_TB$ remaining bits.
\end{itemize}
See Figure \ref{fig:bit-partition} for an illustration. In our implementation, we choose $n\_tile$ to be equal to the logarithm of the warp size~: $n\_tile = 5$.

We also define some notation for dealing with indices~: 
\begin{itemize}
    \item \texttt{stitch\_col(col, TB, row)} forms an index by using \texttt{col} for the $n\_tile$ tile column bits, \texttt{TB} for the $n\_TB$ thread block bits and \texttt{row} for the $n\_tile - n\_over$ remaining tile row bits.
    
    \item \texttt{stitch\_row(col, TB, row)} forms an index by using \texttt{row} for the $n\_tile$ tile row bits, \texttt{TB} for the $n\_TB$ thread block bits and \texttt{col} for the $n\_tile - n\_over$ remaining tile column bits.
    
    \item \texttt{stitch\_tile\_col(col, row)} forms an index as in \texttt{stitch\_col}, but deletes the thread block bits.
    
    \item \texttt{stitch\_tile\_row(col, row)} forms an index as in \texttt{stitch\_row}, but deletes the thread block bits.
\end{itemize}

We show some examples of using these functions for the cyclic shift permutation of Figure \ref{fig:bit-partition}. This permutation shifts the bits of the input index by one position towards the LSB and moves the LSB to the MSB, so that~:
\[
\texttt{cyclic\_shift}(0\texttt{b}11001, 5) = 0\texttt{b}11100
\]
In these examples $n = 10$ and $n\_tile = 5$ (thus $n\_over = 4$ and $n\_TB = 4$), and we follow the usual convention for binary literals of writing the LSB to the right and MSB to the left~:
\begin{Small}
\begin{Verbatim}[commandchars=\\\{\}]
  stitch_col(\red{11010}, 1100, \blue{1}) = 1100\blue{1}\red{11010}
  stitch_tile_col(\red{11010}, \blue{1}) = \blue{1}\red{11010}
  stitch_row(\red{0}, 1011, \blue{00110}) = 1011\blue{00110}\red{0}
  stitch_tile_row(\red{0}, \blue{00110}) = \blue{00110}\red{0}
\end{Verbatim}
\end{Small}

Note that when $n\_over = 0$, \texttt{stitch\_col} and \texttt{stitch\_row} are identical, and \texttt{stitch\_tile\_col} and \texttt{stitch\_tile\_row} are also identical. We refer the reader to the appendix for some intuition on how these stitching functions are translated to CUDA instructions.

Fixing the thread block bits and choosing every possible combination of tile bits defines a single tile~: the input array is thus covered by $2^{n\_TB}$ disjoint tiles. As in the transposition case, we launch one thread block per tile, each of size $2^{n\_tile} * 2^{n\_tile - n\_over}$.

\begin{Small}
\begin{verbatim}
kernel bpc_permutation(int* input, int* output)
{
  shared tile[2^(2*n_tile - n_over)];
  size_t block = blockIdx.x;
  size_t warp = threadIdx.y;
  size_t thread = threadIdx.x;

  // Read the tile.
  tile[stitch_tile_col(thread, warp)] =
      input[stitch_col(thread, TB, warp)];

  // Synchronize
  syncthread();

  // Write the tile
  output[p(stitch_row(warp, TB, thread)) XOR c] = 
      tile[stitch_tile_row(warp, thread)];
}
\end{verbatim}
\end{Small}

This kernel uses only coalesced memory accesses. We can easily see that when reading the input tile each warp reads $2^{n\_tile}$ consecutive elements. This is less clear when writing the output tile. Notice that using $p$ to permute the bits of \texttt{stitch\_row(warp, TB, thread)} moves the bits of \texttt{thread} to the $n\_tile$ lower bits of the index~: each warp thus writes $2^{n\_tile}$ consecutive elements.

\subsection{Avoiding bank conflicts}
\label{subsec:bpc-banks}

The previous kernel solved the coalescing problem, but unfortunately it introduced shared memory bank conflicts, specifically in the second access to the tile in shared memory. 

At this point there are two natural ways of viewing the two-dimensional tile in shared memory~: we could view it as a $2^{n\_tile} * 2^{n\_tile - n\_over}$ matrix, or as a $2^{n\_tile-n\_over} * 2^{n\_tile}$ matrix. We choose the latter option as it yields an easier analysis of bank conflicts. Note that the tile is in general not square~: it can have fewer rows than columns.

\begin{figure}[H]
  \centering
  \includegraphics[width=\linewidth]{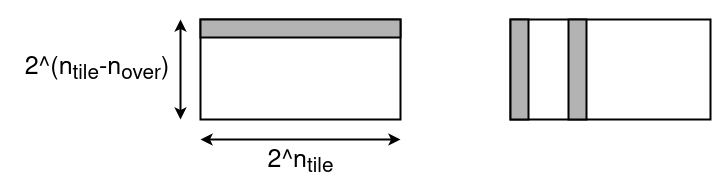}
  \caption{The two-dimensional tile in shared memory when $n\_over = 1$. The shaded region corresponds to the elements accessed by a single warp~: on the left for the first access and on the right for the second access.}
  \label{fig:tile-matrix-accesses}
\end{figure}

We can now analyse both accesses to the tile using this new lens (see Figure \ref{fig:tile-matrix-accesses} for an illustration)~:
\begin{itemize}
    \item In the first access each warp writes a single row in the tile. Since we chose $2^{n\_tile}=32$ this is always bank conflict-free.
    \item In the second access each warp reads $2^{n\_over}$ distinct columns from the tile~: in particular when $n\_over = 0$ each warp reads a single column. Note that the accessed columns are not necessarily evenly spaced.
\end{itemize}

Accessing a matrix column-wise in shared memory results in a bank conflict. In this case, the second access is serialized into $2^{n\_tile - n\_over}$ conflict-free reads, one for each row. 

To fix this conflict we change slightly the way the tile matrix is stored in shared memory~: we shift each row by a given amount to the right. Elements that overflow the end of the row wrap around to the start of the row. More formally, the element at row $i$ and column $j$ is stored at index~:
\begin{small}
$$i * 2^{n\_tile} + (\texttt{shift\_i} + j \texttt{ mod } 2^{n\_tile})$$
\end{small}

We choose the shift for each row depending on the permutation $p$, but note that no matter how we choose the shifts the first access to the tile will always remain conflict-free. We make the following choice~:
\begin{small}
$$\texttt{shift\_i} = \texttt{stitch\_tile\_row}(i, 0)$$
\end{small}
For instance when $n\_over = 0$ we have $\texttt{shift\_i} = i$. We can now analyse the second access again. Each thread accesses the shared memory tile at position $(i, j)$ where~:

\begin{small}
\begin{align*}
  i &= \texttt{stitch\_tile\_row}(\texttt{warp}, \texttt{thread}) \mathbin{/} 2^{n\_tile}\\
    &= \texttt{thread} \mathbin{/} 2^{n\_over}\\
  j &= \texttt{stitch\_tile\_row}(\texttt{warp}, \texttt{thread}) \texttt{ mod } 2^{n\_tile}\\
  &= \texttt{stitch\_tile\_row}(\texttt{warp}, \texttt{thread mod } 2^{n\_over})\\
\end{align*}
\end{small}
This element is in the following bank (modulo $2^{n\_tile}$)~:
\begin{small}
\begin{align*}
  & \texttt{bank}(\texttt{warp}, \texttt{thread}) \\
  &= \texttt{shift\_i} + j \\
  &= \texttt{stitch\_tile\_row}(i, 0) + j \\
  &= \begin{aligned}[t]
    & \texttt{stitch\_tile\_row}(\texttt{thread} \mathbin{/} 2^{n\_over}, 0) \: \mathbin{+} \\
    & \texttt{stitch\_tile\_row}(\texttt{warp}, \texttt{thread mod } 2^{n\_over}) \\ 
  \end{aligned}\\
  &= \begin{aligned}[t]
    & \texttt{stitch\_tile\_row}(\texttt{warp}, 0) \: \mathbin{+} \\
    & \texttt{stitch\_tile\_row}(\texttt{thread} \mathbin{/} 2^{n\_over}, \texttt{thread mod } 2^{n\_over}) \\ 
  \end{aligned}\\
\end{align*}
\end{small}
The final call to \texttt{stitch\_tile\_row} is a bit permutation of \texttt{thread}. This means that in the second access each warp accesses every bank once. The resulting kernel is fully conflict-free.

\subsection{Amortizing index computations}
\label{subsec:bpc-iters}

The running time of the transpose kernel shown in the introduction is almost completely spent on memory operations. This is not the case for more complex permutations (for instance when $n\_over > 0$ or when the tile row bits are not contiguous); the scalar instructions performed by each thread to stitch the bits of input and output indices account for a non-negligeable portion of the running time. We can reduce this overhead by having each thread process $2^{n\_iter}$ input indices instead of only one (typically $n\_iter = 3$).

\begin{figure}[htp]
  \centering
  \begin{subfigure}[b]{0.45\textwidth}
    \centering
    \includegraphics[width=\textwidth]{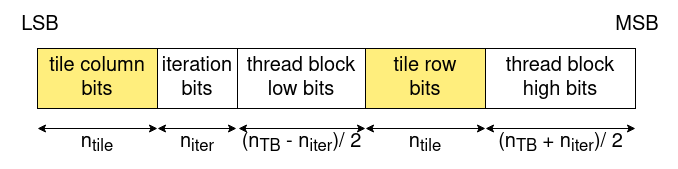}
    \caption{Transpose (for a square matrix).}
  \end{subfigure}
  
  \begin{subfigure}[b]{0.45\textwidth}
    \centering
    \includegraphics[width=\textwidth]{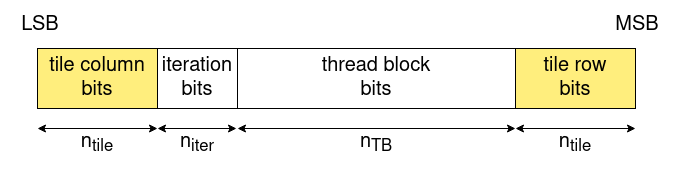}
    \caption{Bit reverse.}
  \end{subfigure}
  
  \caption{Partition of input index bits for different permutations, accounting for the iteration bits. The shaded areas represent the tile bits.}
  \label{fig:bit-partition-iter}
\end{figure}

We modify the partition of input index bits by splitting the thread block bits into two parts~: the lower $n\_iter$ bits become the iteration bits and the upper bits become the new thread block bits (see Figure \ref{fig:bit-partition-iter} for an illustration). The \texttt{stitch\_row} and \texttt{stitch\_col} functions are modified accordingly. 

Each thread block processes $2^{n\_iter}$ tiles~: it reads the tiles sequentially, synchronizes the threads, and writes the tiles sequentially. For instance the read step becomes~:

\begin{Small}
\begin{verbatim}
// Read the tiles.
for (int iter = 0; iter < 2^n_iter; iter++) {
    tiles[iter][stitch_tile_col(thread, warp)] =
        input[stitch_col(thread, iter, TB, warp)];
}
\end{verbatim}
\end{Small}

The advantage of writing the kernel this way is that most index computations can be pulled out of the \texttt{for} loop. Only the parts that depend on \texttt{iter} need remain in the loop (see the appendix for an example). The average amount of scalar instructions per input element is thus greatly reduced.

\section{Implementing BMMC permutations}
\label{sec:bmmc}

\subsection{Tiled BMMCs}
\label{sec:tiled-bmmc}

\begin{figure}[htp]
    \centering
    \includegraphics[width=0.5\linewidth]{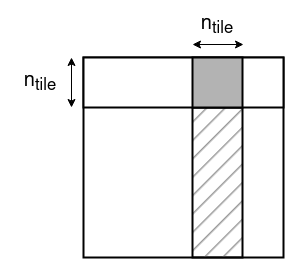}
    \caption{Decomposition of a tiled BMMC. The shaded sub-matrix is invertible and the dashed sub-matrix is equal to $0$. In this example the columns $i_1, \dots, i_{n\_tile}$ are contiguous.}
    \label{fig:tiled-bmmc-condition}
\end{figure}

It is straightforward to extend the kernels developed in the previous section to a class of BMMCs slightly larger than BPCs, namely tiled BMMCs. A tiled BMMC $(A, c)$ is a BMMC corresponding to a permutation that can be implemented using the tiled kernel outlined above. The minimal requirements on the matrix $A$ are that we can find a set of columns $i_1, \dots, i_{n\_tile}$ such that~:
\begin{itemize}
    \item The sub-matrix formed by the first $n\_tile$ rows and the columns $i_1, \dots, i_{n\_tile}$ is invertible.
    \item The sub-matrix formed by the last $n - n\_tile$ rows and the columns $i_1, \dots, i_{n\_tile}$ is equal to $0$.
\end{itemize}
See Figure \ref{fig:tiled-bmmc-condition} for an illustration. Note that a BPC is always a tiled BMMC; in this case the columns $i_1, \dots, i_{n\_tile}$ are exactly the indices of the tile row bits.

When implementing the tiled kernel, the bits of each input index are now partitioned in the following way~:
\begin{itemize}
    \item The tile column bits are the $n\_tile$ lower bits.
    \item The tile row bits are the $n\_tile$ bits $i_1, \dots, i_{n\_tile}$.
\end{itemize}
The tile overlap bits and thread block bits are defined as previously. The only modification we have to make to the kernel is to change the calculation of the output address
$$\texttt{p(stitch\_row(...)) XOR c}$$ 
to use a matrix multiplication instead~:
$$\texttt{A * stitch\_row(...) XOR c}$$

Bank conflicts can now be eliminated in the same way as for BPC permutations. However, the next optimization (amortizing the cost of index computations) cannot be applied to tiled BMMC permutations, as it relies on the sparseness of BPC matrices. We show the exact performance impact in section \ref{sec:results}.

\subsection{Factorizing BMMCs into tiled BMMCs}
\label{sec:bmmc-fact}

The main use case for tiled BMMCs is to provide an implementation for arbitrary BMMC permutations. Using the Lower-Upper (LU) decomposition we show that any BMMC can be factorized into a product of at most two tiled BMMCs. 

Let $(A, c)$ be a BMMC. There exist matrices $U$, $L$ and $P$ such that~:
$$A = U L P$$
where $U$ is an upper triangular matrix, $L$ is a lower triangular matrix and $P$ is a permutation matrix. 

\begin{figure}[htp]
  \centering
  \begin{subfigure}[b]{0.24\linewidth}
    \centering
    \includegraphics[width=\textwidth]{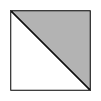}
    \caption*{$U$}
  \end{subfigure}  
  \hfill
  \begin{subfigure}[b]{0.24\linewidth}
    \centering
    \includegraphics[width=\textwidth]{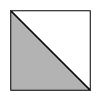}
    \caption*{$L$}
  \end{subfigure}  
  \hfill
  \begin{subfigure}[b]{0.24\linewidth}
    \centering
    \includegraphics[width=\textwidth]{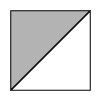}
    \caption*{$U R$}
  \end{subfigure}  
  \hfill
  \begin{subfigure}[b]{0.24\linewidth}
    \centering
    \includegraphics[width=\textwidth]{img/upper_left_shape.drawio.png}
    \caption*{$R L$}
  \end{subfigure}  
  
  \caption{The non-zero entries in each matrix can only occur in the shaded area.}
  \label{fig:UL-shapes}
\end{figure}

Observe that $U$ is the matrix of a tiled BMMC (using the first $n\_tile$ columns) and $P$ is the matrix of a BPC, but $L$ has no such property. We can factorize $A$ in a slightly different way, using the matrix $R$ corresponding to bit-reverse (see section \ref{sec:key-ideas}) such that $R_{ij} = 1$ exactly when $i + j = n-1$ (thus $R^2$ is the identity matrix)~:
$$A = (U R) (R L P)$$
Both factors in this new decomposition are matrices of tiled BMMCs (see Figure \ref{fig:UL-shapes})~:
\begin{itemize}
    \item $U R$ using the columns $n-n\_tile, \dots, n-2, n-1$.
    \item $R L P$ using the columns $p(n-n\_tile), \dots, p(n-2), p(n-1)$.
\end{itemize}

The permutation defined by $(A, c)$ can thus be realized by first permuting using $(RLP, 0)$ and then using $(UR, c)$.

\begin{figure*}[htb]
  \centering
  \includegraphics[width=0.9\textwidth]{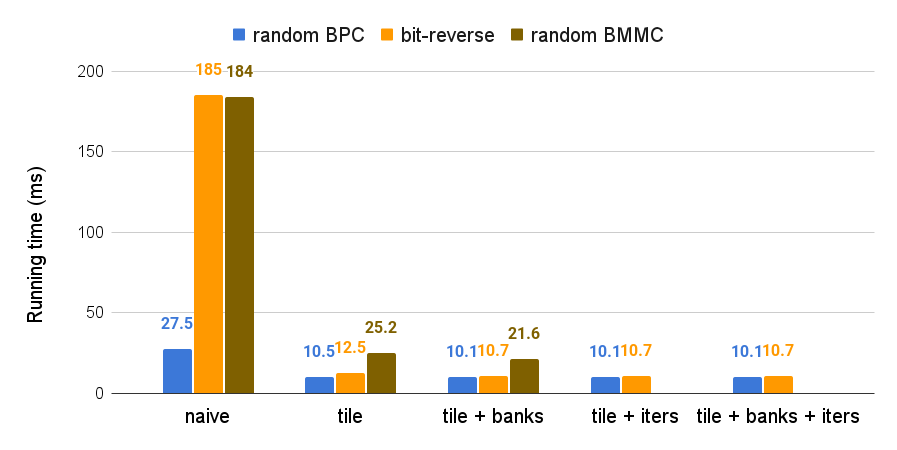}
  \caption{Impact of the different optimizations on running time, for random BPC and BMMC permutations, and for a particular BPC permutation (bit-reversal, the slowest BPC permutation) on arrays of size $2^{30}$. For comparison, the running time of a copy kernel was \SI{9.3}{ms}. The \textit{iters} optimization does not apply to BMMC permutations (see section \ref{sec:tiled-bmmc}).}
  \label{fig:optimization-times}
\end{figure*}
 
\section{Results}
\label{sec:results}

We implemented the kernels outlined above in CUDA~: we use Haskell to generate a CUDA kernel for each permutation. We refer the reader to the appendix for an example of the naive and various optimized BPC permutation kernels.

We used CUDA events to measure the running time of each kernel on a NVIDIA RTX4090 GPU and averaged each measurement across 1000 runs. Unless otherwise noted, all arrays contain 32-bit elements. We report the impact of different optimizations in Figure \ref{fig:optimization-times}~:
\begin{itemize}
  \item The \textit{tile} optimization refers to the tiling optimization described in section \ref{subsec:bpc-coalesce}.
  \item The \textit{banks} optimization refers to the shared memory bank conflict optimization described in section \ref{subsec:bpc-banks}.
  \item The \textit{iters} optimization refers to the iteration optimization described in section \ref{subsec:bpc-iters}. As explained at the end of section \ref{sec:tiled-bmmc} this is only applicable to BPC permutations, not to tiled or arbitrary BMMC permutations.
\end{itemize}

The \textit{tile} optimization yields the largest speedup. For the other two optimizations, we report only the additional speedup when they are added to \textit{tile}.

Our optimized BPC permutation (\textit{tile} + \textit{banks} + \textit{iters}) is about as fast as a simple copy, whereas our optimized BMMC permutation (\textit{tile} + \textit{banks}) is about half as fast as a simple copy. This is because a BMMC permutation is implemented as two tiled kernels and thus does twice the work of a BPC permutation which is implemented as a single tiled kernel. The cost of the binary matrix-vector product performed by each thread in the tiled BMMC kernel accounts for only a few percent of the total running time.

The first column (corresponding to the naive kernels) deserves some explanation. On average, a BPC permutation is much faster than the worst case (corresponding to bit-reversal). This is because a random BPC permutation is likely to have $n\_over > 0$, which means that with the naive kernel each warp writes to only $16$ (when $n\_over = 1$) or even $8$ (when $n\_over = 2$) global memory segments instead of $32$ in the worst case~: the naive kernel is already somewhat coalesced. On the contrary, when choosing a random BMMC permutation and factorizing it as in section \ref{sec:bmmc-fact}, the resulting tiled BMMC permutations almost always have $n\_over$ equal to $0$, meaning that with the naive kernel each warp writes to $32$ global memory segments.

\begin{figure}[htp]
  \centering
  \includegraphics[width=0.9\linewidth]{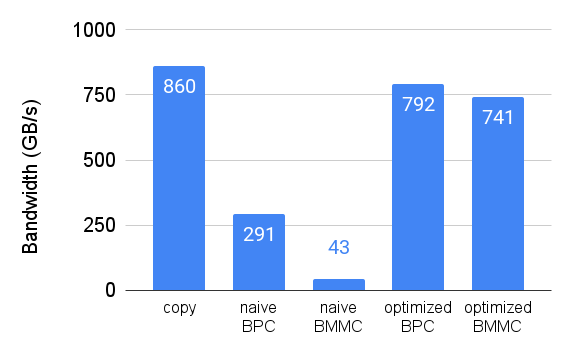}
  \caption{Global memory bandwidth of our kernels (both naive and with all optimizations), measured on arrays of size $2^{30}$. The first column shows that the maximum effective bandwidth of \SI{860}{GB/s} is lower than the maximum theoretical bandwidth, which is \SI{1008}{GB/s} for our GPU.}
  \label{fig:max-bandwith}
\end{figure}

Figure \ref{fig:max-bandwith} shows that our kernels are close to optimal in terms of memory bandwidth~: the optimized BPC and BMMC permutations reach respectively $92\%$ and $86\%$ of the maximum effective bandwidth. Note that memory bandwidth is a measure of how well a memory-bound GPU program uses the memory system and does not directly reflect the program's running time, as the latter also depends on how much data is transferred to and from memory. Recall that the BMMC implementation does twice as much memory transfers as the BPC implementation, which explains why the last two columns are similar although BMMC permutations are twice as slow.

\begin{figure}[htp]
  \centering
  \includegraphics[width=\linewidth]{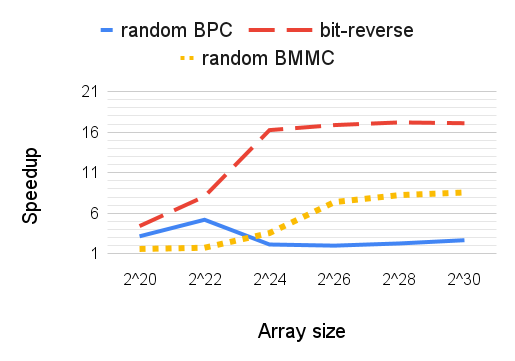}
  \caption{Speedup for different array sizes.}
  \label{fig:speedup-wrt-n}
\end{figure}

Figure \ref{fig:speedup-wrt-n} shows the speedup we obtain using all optimizations compared to the naive version for different array sizes. Compared to Figure \ref{fig:optimization-times}, for arrays of size smaller than $2^{24}$ we get a lower speedup in the random BMMC and bit-reverse case but a higher speedup in the random BPC case (in all cases the speedup is greater than $1$). We do not report data for arrays of size smaller than $2^{20}$~: 
\begin{itemize}
  \item For arrays of size smaller than $2^{20}$, the running time of permutation kernels - both naive and optimized - is only a couple microseconds, which is very close to the GPU clock precision (half a microsecond according to the CUDA Runtime API ~\cite{cuda-runtime-api}, section 6.5 "Event Management").
  \item GPUs need a very large amount of threads to be \textit{saturated}, i.e. to be able to hide global memory latency by switching threads. This is not anymore the case when permuting a single small array~: for instance with the optimized BPC permutation kernel and an array of size $2^{18}$ we would launch $2^{15} = 32768$ threads, which is not enough to saturate the RTX4090 GPU used for benchmarking. 
\end{itemize}

Our current approach for implementing BMMC permutations does have several limitations. We elaborate on the main ones here. Array sizes are restricted to powers of $2$~: we have not yet found a satisfactory way to extend our results to arrays of arbitrary size. We also work in an offline setting, i.e. we assume that the BMMC matrix and complement vector are known at compile time. Extending our approach to work in an online setting would raise some difficulties~:
\begin{itemize}
  \item The decomposition of a BMMC matrix into a product of tiled BMMCs can be a costly operation for large arrays, and is poorly suited to GPUs. 
  \item Implementing the bit-stitching functions used in section \ref{sec:bpc} in an online setting could lead to slowdown due to the additional scalar instructions we would have to generate. While this might not be an issue for BPC permutations since we can use the optimization outlined in section \ref{subsec:bpc-iters} to alleviate the cost of scalar instructions, this would certainly result in at least a minor slowdown for arbitrary tiled BMMC permutations.
\end{itemize}

\begin{figure}[htp]
    \centering
    \includegraphics[width=\linewidth]{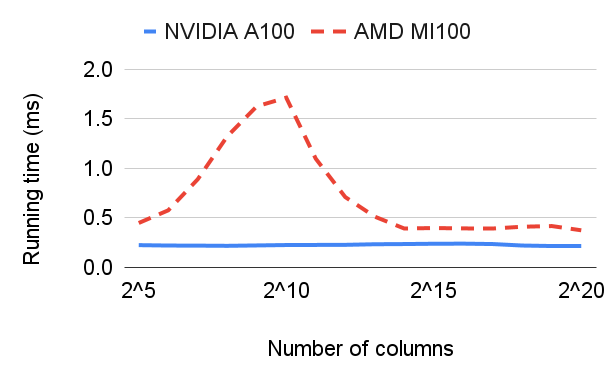}
    \caption{Running time of an optimized transpose kernel on an AMD and NVIDIA GPU, for matrices of various sizes. We keep the number of elements constant equal to $2^{25}$ and vary the number of columns (always a power of 2).}
    \label{fig:futhark-transpose-graph}
\end{figure}

All the measurements in this article were performed on a NVIDIA RTX4090 GPU. We could not reproduce them on an AMD GPU~: we ran into some unexpected slowdowns related to global memory. Despite being fully coalesced, the running time of our tiled permutation kernels depended heavily on the given BPC or BMMC matrix. This can be reproduced even with a kernel as simple as a tiled transpose~: see Figure \ref{fig:futhark-transpose-graph} for an example using the Futhark transpose kernel. This phenomenon only occurs when array sizes are powers of two, and as such is not an issue for most Futhark programs, but is an issue for the algorithms in this paper. 

There seem to be differences in the memory architecture between AMD and NVIDIA. Our guess is that they have a different address mapping scheme and that our kernels trigger global memory bank conflicts on AMD cards, however we have not been able to prove or disprove this intuition and are open to suggestions. We refer the reader to ~\cite{bim-address-mapping} for a discussion on GPU address mapping schemes that coincidentally makes use of BMMCs.

\section{Application to the parm combinator}
\label{sec:combinators}

\subsection{Using the parm combinator}
\label{subsec:parm}

As a use case of BMMC permutations we describe how they can be used to implement a high level combinator called \texttt{parm}. This is not the only useful combinator that is related to BMMCs~: other examples are outside the scope of this paper, but we do plan on studying these combinators further in future work. We refer the reader to ~\cite{obsidian-paper} for another paper using similar combinators.

Let us remind how the \texttt{parm} combinator works~: it takes as input an array xs of size $2^n$, an $n$-bit binary mask and a function f that maps arrays of size $2^{n-1}$ to arrays of size $2^{n-1}$. The input array xs is partitioned into two sub-arrays xs0 and xs1 depending on the mask as follows (see Figures \ref{fig:parm-examples} and \ref{fig:parm-matrix-example})~:
\[
\texttt{sub-array}(\texttt{i}) = \begin{cases}
  \texttt{xs0} & \texttt{ if i } * \texttt{ mask} = 0 \\
  \texttt{xs1} & \texttt{ if i } * \texttt{ mask} = 1 \\
\end{cases}
\]
Where i is the index of the given element in xs and \texttt{*} denotes the dot product in $F_2$. We then apply f to each sub-array and stitch them back together in exactly the same way.

We now show how to use \texttt{parm} to implement a simple sorting network, inspired by Batcher's bitonic sorting network ~\cite{batcher-sorting-networks} and the balanced periodic merger~\cite{balanced-sorting-network}. There has been previous effort to generate efficient GPU code for such networks~: see~\cite{obsidian-paper} for an approach that focusses on small networks operating on arrays that fit in shared memory.

The network we study in this example is a variant of merge sort: the elements at even and odd indices are sorted separately before being merged. The following function sorts its input \texttt{xs} of size $2^n$~:

\begin{Small}
\begin{verbatim}
sort 0 xs = xs
sort n xs = let ys = parm 1 (sort (n-1)) xs
            in merge n ys
\end{verbatim}
\end{Small}

\begin{figure}[htp]
    \centering
    \includegraphics[width=0.7\linewidth, height=0.6\linewidth]{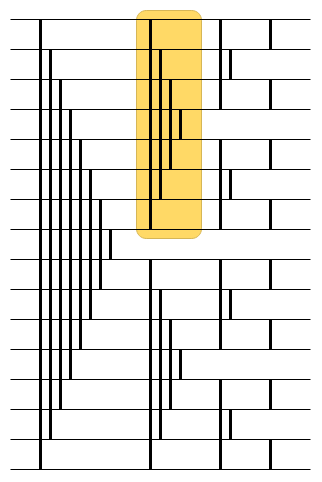}
    \caption{A 16 input balanced periodic merger. The shaded region corresponds to an 8 input V-column.}
    \label{fig:merging-network}
\end{figure}

The merge function takes as input an array in which the two sub-arrays formed by the elements at even and odd indices are sorted and produces a sorted output. We choose to use a balanced periodic merger~: Figure~\ref{fig:merging-network} illustrates the merging network. Data flows from left to right along the $16$ horizontal lines. The vertical lines operate on two inputs and place the minimum on the top and the maximum on the bottom. Here is the corresponding pseudocode~:

\begin{Small}
\begin{verbatim}
merge 0 xs = xs
merge n xs = let ys = vcolumn n xs
             in parm (2^(n-1)) (merge (n-1)) ys
\end{verbatim}
\end{Small}

The \texttt{vcolumn} function in turn builds a single V-shaped column with $2^n$ inputs in the merging network. This can be accomplished by simply interleaving two half-size V-columns using a mask equal to $3 = 0\texttt{b}11$ (see also Figure~\ref{fig:parm-examples}).

\begin{Small}
\begin{verbatim}
vcolumn 0 xs = xs
vcolumn 1 [x1, x2] = 
  if x1 <= x2 then [x1, x2] else [x2, x1]
vcolumn n xs = parm 3 (vcolumn (n-1)) xs
\end{verbatim}
\end{Small}

The \texttt{parm} combinator shines here because it allows the programmer to specify the sorting network in a declarative style, leaving many opportunities for the compiler to optimize the program (in this case using BMMCs to permute arrays and obtain coalesced memory accesses). 

\subsection{Compiling parm using BMMC permutations}
\label{subsec:parm-bmmc}

While the above example shows the expressiveness of \texttt{parm}, a straightforward implementation - in which the function f we apply to each sub-array reads its inputs directly from \texttt{xs} and writes directly to the output array - is not suited to GPUs. To gain some intuition on why consider the case where f makes only fully coalesced reads and writes. For most masks (think for instance of \texttt{mask $= 1$}) the resulting function \texttt{parm mask f} will not make fully coalesced accesses, and in fact will require twice as much memory transactions as a coalesced version would. Now take into account that \texttt{parm} is often nested many times (as in the sorting network example) and we loose all coalescing.

Our solution for compiling \texttt{parm mask f xs} while retaining coalescing is to first permute the array \texttt{xs} such that the two subarrays \texttt{xs0} and \texttt{xs1} form the first and second half of the resulting array, apply f to each half and then permute the array back. When applying f, the two sub-arrays are contiguous in memory~: any coalescing behaviour of f will therefore be retained. Permuting the array twice (before and after applying f) of course adds some overhead~: however these permutations are in fact BMMC permutations, allowing for an efficient implementation. 

We now explain how to construct a matrix $A$ such that~:
\[
\texttt{parm} \: m \: f = \texttt{bmmc} \: (A^{-1}, 0) \circ \texttt{parm} \: 2^{n-1} \: f \circ \texttt{bmmc} \: (A, 0)
\]
Permuting \texttt{xs} using the BMMC $(A, 0)$ should put \texttt{xs0} into the first half and \texttt{xs1} into the second half, while preserving the order of elements within each sub-array. More formally, an element at index $x$ in \texttt{xs} should have the index $y$ in the result such that~:
\begin{itemize}
  \item[] $y_{0..n-2} = \texttt{sub-index}(x)$
  \item[] $y_{n-1} = \texttt{sub-array}(x)$ 
\end{itemize}
Where \texttt{sub-array(x)} is equal to $0$ if $x$ is in the first sub-array and $1$ otherwise, and \texttt{sub-index(x)} is the new index of the element at position $x$ in its sub-array (see Figure \ref{fig:parm-matrix-example} for an example).

\begin{figure}[htp]
\centering
  \begin{subfigure}[c]{0.7\linewidth}
    \centering
    \begin{tabular}{|c|c|c|}
      \hline
      index $x$ & sub-array & sub-index \\
      \hline
      0 & 0 & 0 \\
      1 & 0 & 1 \\
      2 & 1 & 0 \\
      3 & 1 & 1 \\
      4 & 1 & 2 \\
      5 & 1 & 3 \\
      6 & 0 & 2 \\
      7 & 0 & 3 \\
      \hline
    \end{tabular}
    \caption{The \texttt{sub-array} and \texttt{sub-index} functions.}
  \end{subfigure}
  \hfill
  \begin{subfigure}[c]{0.29\linewidth}
    \centering
    \[
    \begin{bmatrix}
    1 & 0 & 0 \\
    0 & 0 & 1 \\
    0 & 1 & 1 \\
    \end{bmatrix}
    \]
    \caption{The matrix $A$.}
  \end{subfigure}
  \caption{Constructing the matrix $A$ for an $8$ element array and a mask equal to $6 = 0\texttt{b}110$. In this case $\texttt{lsb(mask)} = 1$.}
  \label{fig:parm-matrix-example}
\end{figure}

Notice that \texttt{sub-array(x)} is simply equal to $i * \texttt{mask}$. Finding an expression for \texttt{sub-index(x)} is slightly harder. It turns out that it is sufficient to remove the bit at index \texttt{lsb(mask)} from $x$, where \texttt{lsb(mask)} is the index of the least significant bit of the mask. The reader is invited to check this fact in Figure \ref{fig:parm-matrix-example}. This yields the following relation between $x$ and $y$ from which it is straightforward to construct the matrix $A$ (a similar formula can be derived for $A^{-1}$)~:
\[ 
y_i = \begin{cases}
  x_i & \texttt{ if } i < \texttt{lsb(mask)} \\
  x_{i+1} & \texttt{ if } \texttt{lsb(mask)} \leq i < n-1 \\
  x * \texttt{mask} & \texttt{ if } i = n-1 \\
\end{cases}
\]

It should be noted that \texttt{parm} and \texttt{bmmc} give rise to a rich set of rewrite rules that allow us to reduce the number of BMMC permutations performed in most cases, especially when nesting applications of \texttt{parm}. 

\section{Related Work}

BMMCs were first studied by Cormen in the setting of the parallel disk I/O model introduced by Vitter and Schriver~\cite{vitter-disk-model}. This model consists in a processor (or multiprocessor) connected to several storage devices which can be accessed in parallel, and places an emphasis on the memory system rather than on the processor. Performance in this model is measured in terms of I/O accesses. Cormen showed how to perform BMMC and BPC permutations for large on-disk arrays and proved optimality results for his implementations in terms of number of memory accesses~\cite{cormen-bmmc-bounds, cormen-bmmc-implementation}. This inspired our current work, which tackles the same problem but in the context of GPUs, for which memory access performance is just as important as in the context of parallel disk I/O.

There have been previous attempts at performing permutations efficiently on GPUs~: Kasagi et al.~\cite{cuda-offline-permutations} show how to implement arbitrary permutations (also in an offline setting) in a fully coalesced and bank-conflict free manner, and additionally provide specialized kernels for specific permutations such as transpose or bit-reverse. Their method has similar theoretical guarantees in terms of bandwidth as ours, but they use 5 kernels per permutation whereas we use only one and two kernels for BPC and BMMC permutations respectively. The result is that while our permutations reach roughly 50\% (for BMMCs) and 100\% (for BPCs) of the speed of a copy, their fastest algorithm is 5 times slower than a copy. Kasagi's method is additionally limited by shared memory size~: for an input array of $N$ elements, it requires that $\sqrt{N}$ elements can fit in shared memory, which is typically only a few kilobytes on modern GPUs. Their method can thus only handle input arrays of up to roughly $2^{24}$ 32-bit elements.

More recently, BMMCs have been used to design GPU address mapping schemes~\cite{bim-address-mapping}. To the programmer, GPU global memory is presented as a single contiguous block of memory. The translation between a memory address and actual hardware parameters (involving a bank index, channel index and so on) is handled by a so-called address mapping scheme. Liu et al. represent this mapping using a BMMC mapping~: in essence, they implement a fixed BMMC permutation directly in GPU hardware.

\section{Conclusions}

We have shown an efficient CUDA implementation of BMMC permutations, a class that includes many interesting permutations. The benchmark results are promising, especially for BPC permutations which are basically as fast as they can get, reaching upwards of $90\%$ of the maximum effective bandwidth. 

We also explained how inserting BMMC permutations in GPU code at the right places can allow for fully coalesced memory accesses. In some sense, this generalizes an optimization present in the Futhark compiler in which multidimensional arrays are automatically transposed in memory to create opportunities for coalescing when possible (\cite{futhark-pldi} section 5.2 "Optimizing Locality of Reference"). In both cases this does create a tradeoff between the speedup from coalescing and the slowdown from executing additional permutations. Our aim moving forward is to implement \texttt{parm} and several related combinators in the Futhark compiler to measure the net gains of this tradeoff. These combinators come with a rich fusion algebra which should permit further optimizations.

\section*{Acknowledgement}
This research was funded by a Swedish Research Council grant "An algebra of array combinators and its applications", proj. no. 2021-05491. We would also like to thank Troels Henriksen for providing the data for Figure \ref{fig:futhark-transpose-graph}.

\section*{Appendix~: generated CUDA kernels}

This appendix shows the complete CUDA kernels generated for the bit-reverse permutation. For all kernels in this section, the parameters are as follows~:
\begin{center}
\begin{tabular}{ccc}
  n = 15 & n\_tile = 5 & n\_over = 0 \\
\end{tabular}
\end{center}

The number of scalar instructions in these kernels might be higher than expected~: we deliberately do not use CUDA intrinsic functions such as \texttt{\_\_brev()} to speed up index computations as this approach would not work for arbitrary bit permutations. We do however perform a simple optimization to reduce the instruction count. When setting bits $i_0 < \dots < i_k$ in a destination variable using bits $j_0 < \dots < j_k$ respectively in an input variable, if the offsets $i_1 - i_0, \dots, i_k - i_{k-1}$ are equal to the offsets $j_1 - j_0, \dots, j_k - j_{k-1}$, we set all the bits in a single operation (corresponding to a single line in the kernels below). We measured the impact of this optimization and found that on average it reduced by $50\%$ the number of scalar instructions that were generated.

Here is the naive kernel with no tiling~:
\begin{Small}
\begin{verbatim}
__global__ void bit_reverse_naive(
    const int* input, int* output) {
  size_t in_addr = blockIdx.x * blockDim.x + threadIdx.x;
    
  // Compute the output address
  size_t out_addr = 0;
  out_addr |= (in_addr & 1ULL) << 14;
  out_addr |= (in_addr & 2ULL) << 12;
  out_addr |= (in_addr & 4ULL) << 10;
  out_addr |= (in_addr & 8ULL) << 8;
  out_addr |= (in_addr & 16ULL) << 6;
  out_addr |= (in_addr & 32ULL) << 4;
  out_addr |= (in_addr & 64ULL) << 2;
  out_addr |= in_addr & 128ULL;
  out_addr |= (in_addr & 256ULL) >> 2;
  out_addr |= (in_addr & 512ULL) >> 4;
  out_addr |= (in_addr & 1024ULL) >> 6;
  out_addr |= (in_addr & 2048ULL) >> 8;
  out_addr |= (in_addr & 4096ULL) >> 10;
  out_addr |= (in_addr & 8192ULL) >> 12;
  out_addr |= (in_addr & 16384ULL) >> 14;
  output[out_addr] = input[in_addr];
}
\end{verbatim}
\end{Small}

Here is the tiled kernel~:
\begin{Small}
\begin{verbatim}
__global__ void bit_reverse_tiled(
    const int* input, int* output) {
  __shared__ int tile[1024];
  size_t block = blockIdx.x;
  size_t warp = threadIdx.y;
  size_t thread = threadIdx.x;
  
  // Read the input tile
  size_t in_addr = 0;
  size_t itile_addr = 0;
  in_addr |= (block & 31ULL) << 5;
  in_addr |= thread & 31ULL;
  in_addr |= (warp & 31ULL) << 10;
  itile_addr |= thread & 31ULL;
  itile_addr |= (warp & 31ULL) << 5;
  tile[itile_addr] = input[in_addr];
  
  // Synchronize
  __syncthreads();
    
  // Write the output tile
  size_t out_addr = 0;
  size_t otile_addr = 0;
  out_addr |= (block & 1ULL) << 9;
  out_addr |= (block & 2ULL) << 7;
  out_addr |= (block & 4ULL) << 5;
  out_addr |= (block & 8ULL) << 3;
  out_addr |= (block & 16ULL) << 1;
  out_addr |= (thread & 1ULL) << 4;
  out_addr |= (thread & 2ULL) << 2;
  out_addr |= thread & 4ULL;
  out_addr |= (thread & 8ULL) >> 2;
  out_addr |= (thread & 16ULL) >> 4;
  out_addr |= (warp & 1ULL) << 14;
  out_addr |= (warp & 2ULL) << 12;
  out_addr |= (warp & 4ULL) << 10;
  out_addr |= (warp & 8ULL) << 8;
  out_addr |= (warp & 16ULL) << 6;
  otile_addr |= (thread & 31ULL) << 5;
  otile_addr |= warp & 31ULL;
  output[out_addr] = tile[otile_addr];
}
\end{verbatim}
\end{Small}

Here is the tiled kernel, bank-conflict free~:
\begin{Small}
\begin{verbatim}
__global__ void bit_reverse_banks(
    const int* input, int* output) {
  __shared__ int tile[1024];
  size_t block = blockIdx.x;
  size_t warp = threadIdx.y;
  size_t thread = threadIdx.x;
  
  // Read the input tile
  size_t in_addr = 0;
  size_t itile_addr = 0;
  size_t ishift = 0;
  in_addr |= (block & 31ULL) << 5;
  in_addr |= thread & 31ULL;
  in_addr |= (warp & 31ULL) << 10;
  itile_addr |= thread & 31ULL;
  itile_addr |= (warp & 31ULL) << 5;
  ishift |= (itile_addr & 992ULL) >> 5;
  tile[(itile_addr & 992) + 
    ((ishift + itile_addr) & 31)] = 
      input[in_addr];
  
  // Synchronize
  __syncthreads();
    
  // Write the output tile
  size_t out_addr = 0;
  size_t otile_addr = 0;
  size_t oshift = 0;
  out_addr |= (block & 1ULL) << 9;
  out_addr |= (block & 2ULL) << 7;
  out_addr |= (block & 4ULL) << 5;
  out_addr |= (block & 8ULL) << 3;
  out_addr |= (block & 16ULL) << 1;
  out_addr |= (thread & 1ULL) << 4;
  out_addr |= (thread & 2ULL) << 2;
  out_addr |= thread & 4ULL;
  out_addr |= (thread & 8ULL) >> 2;
  out_addr |= (thread & 16ULL) >> 4;
  out_addr |= (warp & 1ULL) << 14;
  out_addr |= (warp & 2ULL) << 12;
  out_addr |= (warp & 4ULL) << 10;
  out_addr |= (warp & 8ULL) << 8;
  out_addr |= (warp & 16ULL) << 6;
  otile_addr |= (thread & 31ULL) << 5;
  otile_addr |= warp & 31ULL;
  oshift |= (otile_addr & 992ULL) >> 5;
  output[out_addr] = 
    tile[(otile_addr & 992) + 
      ((oshift + otile_addr) & 31)];
}
\end{verbatim}
\end{Small}

Here is the tiled kernel, using iterations (but susceptible to bank conflicts). We choose $n\_iter = 3$ for this example~:
\begin{Small}
\begin{verbatim}
__global__ void bit_reverse_iters(
    const int* input, int* output) {
  __shared__ int tile[8192];
  size_t block = blockIdx.x;
  size_t warp = threadIdx.y;
  size_t thread = threadIdx.x;
  
  // Read the input tiles
  size_t in_addr = 0;
  size_t itile_addr = 0;
  itile_addr |= thread & 31ULL;
  itile_addr |= (warp & 31ULL) << 5;
  in_addr |= (block & 3ULL) << 8;
  in_addr |= thread & 31ULL;
  in_addr |= (warp & 31ULL) << 10;
  for (size_t iter = 0; iter < 8; iter++) {
      in_addr &= ~224ULL;
      in_addr |= (iter & 7ULL) << 5;
      tile[(iter << 10) + itile_addr] = 
        input[in_addr];
  }
  
  // Synchronize
  __syncthreads();
  
  // Write the output tiles
  size_t out_addr = 0;
  size_t otile_addr = 0;
  otile_addr |= (thread & 31ULL) << 5;
  otile_addr |= warp & 31ULL;
  out_addr |= (block & 1ULL) << 6;
  out_addr |= (block & 2ULL) << 4;
  out_addr |= (thread & 1ULL) << 4;
  out_addr |= (thread & 2ULL) << 2;
  out_addr |= thread & 4ULL;
  out_addr |= (thread & 8ULL) >> 2;
  out_addr |= (thread & 16ULL) >> 4;
  out_addr |= (warp & 1ULL) << 14;
  out_addr |= (warp & 2ULL) << 12;
  out_addr |= (warp & 4ULL) << 10;
  out_addr |= (warp & 8ULL) << 8;
  out_addr |= (warp & 16ULL) << 6;
  for (size_t iter = 0; iter < 8; iter++) {
      out_addr &= ~896ULL;
      out_addr |= (iter & 1ULL) << 9;
      out_addr |= (iter & 2ULL) << 7;
      out_addr |= (iter & 4ULL) << 5;
      output[out_addr] = 
        tile[(iter << 10) + otile_addr];
  }
}
\end{verbatim}
\end{Small}

\newpage

\bibliography{biblio}

\end{document}